\documentclass[aps,pre,twocolumn,showpacs,superscriptaddress,groupedaddress]{revtex4}  % for review and submission
\usepackage{graphicx}  % needed for figures
\usepackage{dcolumn}   % needed for some tables
\usepackage{bm}        % for math
\usepackage{amssymb}   % for math
\usepackage{amsmath}
\usepackage{latexsym}
\usepackage{verbatim}
\usepackage{amsfonts}
\usepackage{color}
\begin{document}

%\widetext
%\leftline{Version xx as of \today}
%\leftline{Primary authors: Joe E. Physics}
%\leftline{To be submitted to (PRL, PRD-RC, PRD, PLB; choose one.)}
%\leftline{Comment to {\tt d0-run2eb-nnn@fnal.gov} by xxx, yyy}
%\centerline{\em D\O\ INTERNAL DOCUMENT -- NOT FOR PUBLIC DISTRIBUTION}

\title{Percolation of networks with directed dependency links}

\author{Dunbiao Niu}
\affiliation{School of Data and Computer Science, 
Sun Yat-sen University, Guangzhou 510006, China}
\affiliation{School of Mathematics, Southwest Jiaotong University,
 Chengdu 610031, China}
\author{Xin Yuan} \affiliation{Center for Polymer Studies and Department
 of Physics, Boston University, Boston, Massachusetts 02215 USA}
\author{Minhui Du}\affiliation{School of Data and Computer Science, 
Sun Yat-sen University, Guangzhou 510006, China}
\author{H. Eugene Stanley}\affiliation{Center for Polymer Studies and 
Department of Physics, Boston University, Boston, Massachusetts 02215 USA}
\author{Yanqing Hu\footnote{Corresponding author. Email: yanqing.hu.sc@gmail.com}}
\affiliation{School of Data and Computer Science, 
Sun Yat-sen University, Guangzhou 510006, China}
\affiliation{Big Data Research Center, 
University of Electronic Science and Technology of China, Chengdu 611731, China}
\date{\today}

\begin{abstract}
The self-consistent probabilistic approach has proven itself powerful in
studying the percolation behavior of interdependent or multiplex networks without tracking
the percolation process through each cascading step. In order to
understand how directed dependency links impact criticality, we employ
this approach to study the percolation properties of networks with both
undirected connectivity links and directed dependency links. We find
that when a random network with a given degree distribution undergoes a
second-order phase transition, the critical point and the unstable
regime surrounding the second-order phase transition regime are
determined by the proportion of nodes that do not depend on any other nodes. 
Moreover, we also find that the triple point and the boundary between 
first- and second-order transitions are determined by the proportion of nodes that depend on no more
than one node. This implies that it is maybe general for multiplex network systems, 
some important properties of phase transitions can be determined only by a few parameters.
We illustrate our findings using Erd\H{o}s-R\'{e}nyi (ER) networks.
\end{abstract}


\maketitle
\section{Introduction}
Complex networks science has become an effective tool for modeling 
complex systems. It treats system entities as nodes and the mutually supporting or cooperating
 relations between the entities as connectivity links \cite{watts1998collective,albert2000error,cohen2000resilience,callaway2000network,newman2002spread,newman2003structure,rosato2008modelling,arenas2008synchronization,reis2014avoiding,cohen2010complex,
newman2010networks,hu2011possible}. In many systems, nodes that survive and fail together 
form dependency groups through dependency links. Dependency links denote the damaging or destructive relations among entities \cite{buldyrev2010catastrophic,JianxiPRE2013,hu2013percolation,GaoGaoPRE2012,liu2016breakdown,havlin2015percolation,XinPhysRevE2015,hu2011percolation,JianxiPRL2011}. 
Compared to ordinary networks \cite{newman2002spread,cohen2010complex,newman2003structure},
networks with dependency groups or links are more vulnerable and subject to catastrophic collapse \cite{parshani2011critical,bashan2011percolation}.
The previous works have studied the network system in which the dependency groups, with sizes either fixed at two \cite{parshani2011critical} or characterized by different classic distributions \cite{bashan2011percolation},
are formed through undirected dependency links. The outcome when the 
dependency links are {\it directed}, however, is more general. For 
example, in a financial network where each company has trading and 
sales connections (connectivity links) with other companies,  the 
connections enable the companies to interact with others and function
together as a global financial market, and companies that belong to 
the same corporate group strongly depend on the parent company 
(i.e. there are directed dependency links), but the reverse is not true 
\cite{li2014ranking}. Another example is in a social network in which 
people (followers) follow trends set by celebrities (pioneers), e.g., popular
 singers and actors but the reverse is not true \cite{hu2014conditions}.

\begin{figure}
\includegraphics[width=0.48\textwidth, angle = 0]{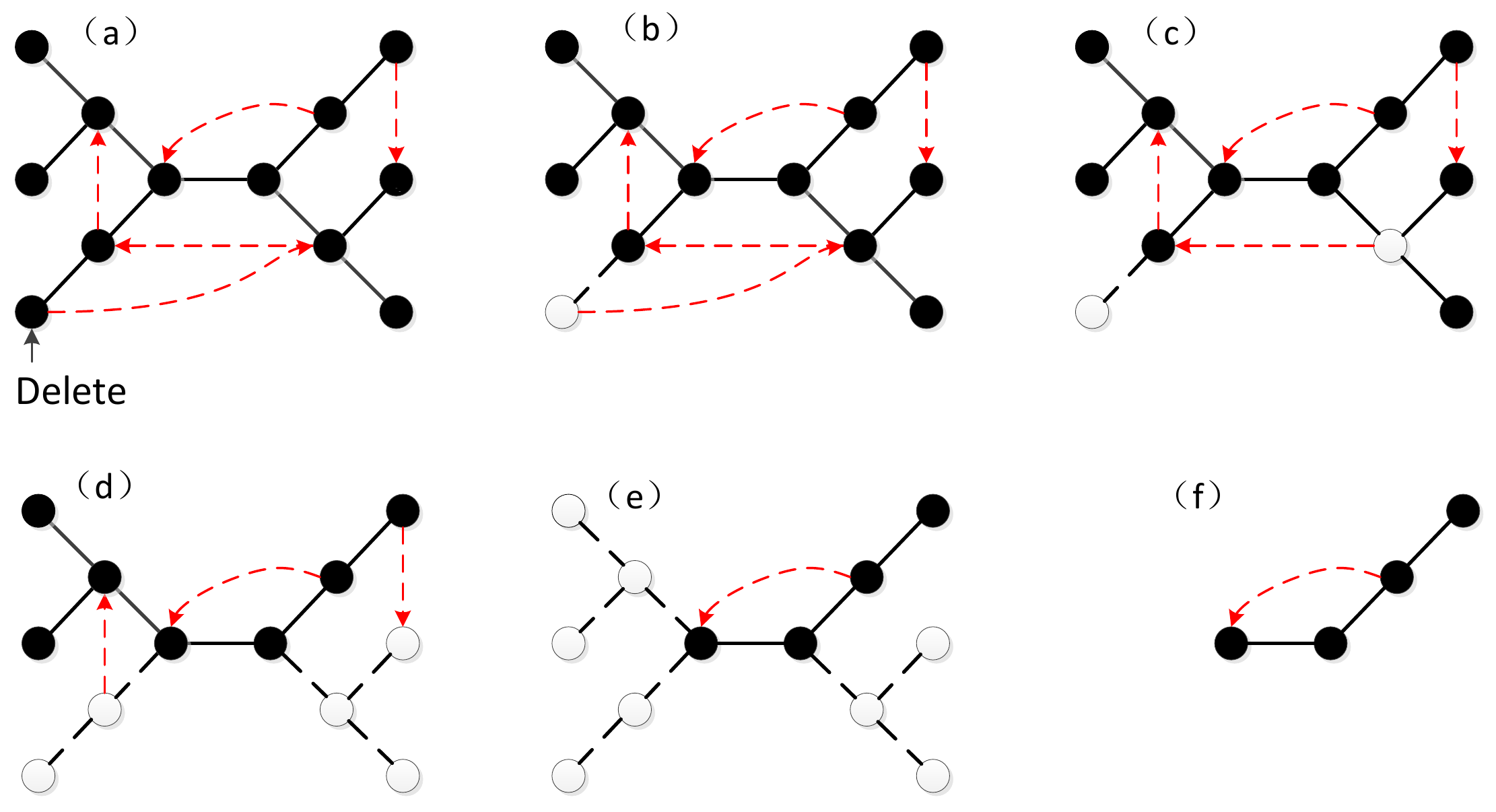}
\caption{\label{Fig1}(Color online) Demonstration of the synergy
between the percolation process and the  dependency process that 
leads to a cascade of failures. The network contains two types of links: 
connectivity links (solid black lines) and directed dependency links 
(dashed red arrows). (a)$\rightarrow$(b) Initial failure: a random node
 is removed. (c)$\rightarrow$(e) Synergy between percolation process 
 and dependency process: nodes cut-off from the giant component or 
 depending on failed nodes are removed. (f) Steady state: the surviving giant 
component contains four nodes. }
\end{figure}  

We use a self-consistent probabilistic framework \cite{MendosPRLKcore2006,BianconiPRE2014,BaxterPRL2012,feng2015simplified} 
to study the percolation phase transitions in a random network $A$
with both connectivity and directed dependency links. Randomly 
removing a fraction $1-p$ of nodes in network $A$ causes (i)
 connectivity links to be disconnected, causing some nodes and 
clusters to fail due to the disconnection to the network giant 
component (percolation process), and (ii) failing nodes to make 
their dependent nodes to also fail even though they are still connected 
to the network giant component via connectivity links (dependency process).
Thus, the removal of nodes in the percolation process leads to the 
failure of dependent nodes in the dependency process, which in turn 
initiates a new percolation process, which further sets off a dependency 
process, and so on. We show that this synergy between the percolation
process and the dependency process leads to a cascade of failures that
 continues until no further nodes fail (See Fig.~\ref{Fig1}).

To fully capture the structure of network $A$,  we 
introduce the degree distribution $P(k)$ and, in addition, the directed 
dependency degree distribution $Q(k_o)$, which is the probability that a 
randomly chosen node has $k_o$ directed dependency links connecting to
$k_o$ nodes which are supporting this chosen node.  In our model, 
when $i$ depends on $k_o$ nodes, we assume that if any one of these 
$k_o$ nodes fails, node $i$ will fail too (see Fig.~\ref{Fig1}).
Usually, this kind of multiplex has both first- 
and second-order phase transitions \cite{parshani2011critical,bashan2011percolation}.
Here we find that $Q(k_o)$ strongly affects the robustness of network $A$. 
Specifically, the percolation threshold $p_c^{II}$, at which network $A$ 
disintegrates in a form of second-order phase transition, is determined solely 
by $Q(0)$ for a given $Q(k_o)$, and $Q(0)+Q(1)$ characterizes the boundary 
between the first-order phase transition and the second-order phase transition regime.

This paper is organized as the follows. In Sec.II we introduce the general 
framework and develop the analytic formulae to solve the influence of 
$Q(k_o)$ on the percolation properties of a random network. In Sec.III, we 
demonstrate these influences using an ER network. 

\section{GENERAL FRAMEWORK}
For a random network $A$ of size $N$ with both connectivity links and 
directed dependency links (see Fig.~\ref{Fig1}(a)), as in  Ref.~\cite{newman2002spread}, 
we introduce the generating function $G_0(z)$ of the degree distribution $P(k)$,
\begin{equation}
G_0(z)=\sum_{k}P(k)z^k.  \label{G_0}
\end{equation}
Analogously, we have the generating function of the related branching processes 
\cite{newman2002spread}, 
\begin{equation}
G_1(z)=\frac{G_0^{'}(z)}{G_0^{'}(1)}=\sum_{k}\frac{kP(k)}{\left\langle k \right\rangle}z^{k-1}. \label{G_1}
\end{equation}
Similarly, we introduce the generating function for the directed dependency 
degree distribution $Q(k_o)$ as 
\begin{equation}
D(z)=\sum_{k_o} Q(k_o)z^{k_o}.  \label{Q}
\end{equation}
We designate $h(s)$ the probability distribution of the number of 
nodes approachable along the directed dependency links starting 
from a randomly chosen node in network $A$. This allows us to write 
the generating function  $H(z)$ for $h(s)$, i.e.,
\begin{equation}
H(z)=\sum_{s}h(s)z^s. \label{H}
\end{equation}
 According to Ref.\cite{cohenpercolationdirected},
$H(z)$ also satisfies a self-consistent condition of the form
\begin{equation} \label{H(x)}
H(z)=z \cdot D(H(z)).
\end{equation}

A random removal of a fraction $1-p$ of nodes triggers a cascade 
of failures. When no more nodes fail, network $A$ reaches its final 
steady state. At this steady state, we use the probabilistic approach 
\cite{feng2015simplified} and {\it define $x$ to be the probability that a 
randomly chosen connectivity link leads to the giant component at one 
of its ends}. If we randomly choose a connectivity link $l$ and find an 
arbitrary node $n$ by following $l$ in an arbitrary direction, the probability 
that node $n$ has degree $k$ is 
\begin{equation}
\frac{kP(k)}{\sum_k kP(k)}=\frac{kP(k)}{\left\langle k \right\rangle}.
\end{equation}
For node $n$,  the root of a directed cluster of size $s$, to be part of 
the giant component, at least one of its other $k-1$ out-going connectivity 
links (other than the link first chosen) leads to the giant component, 
provided that every other $s-1$ node is also in the giant component
because the disconnection of any one of these $s-1$ nodes to 
the giant component will cause node $n$ to lose support and fail. 
Computing this probability, we can write out the self-consistent equation for $x$ as
\begin{eqnarray}\label{x1}
\nonumber x&=&p\{\sum_{k}\frac{kP(k)}{\left\langle k \right\rangle}[1-(1-x)^{k-1}]\}\times \\
&& \sum_{s} \{h(s)\{p\sum_{k}P(k)[1-(1-x)^{k}]\}^{s-1}\},  \label{x_loose}
\end{eqnarray}
where $p$ is the probability that a node survives the initial removal 
process, $1-(1-x)^{k-1}$ is the probability that at 
least one of the other $k-1$ connectivity links of node $n$ leads 
to the giant component, $h(s)$ is the probability
that node $n$ is the root of a directed cluster of size $s$, and
$\{p\sum_{k}P(k)[1-(1-x)^{k}]\}^{s-1}$ is
the probability that every other $s-1$ node in the directed cluster 
supporting node $n$ is also in the giant component.  Using 
the generating functions defined in Eqs.~(\ref{G_0}), (\ref{G_1}) and 
(\ref{H}), we transform Eq.~(\ref{x_loose}) into the compact form
\begin{equation}
x=\frac{1-G_1(1-x)}{1-G_0(1-x)}\cdot H(p[1-G_0(1-x)]). \label{x_compact}
\end{equation}
which, by viewing $p[1-G_0(1-x)]$ as 
a whole and using the property of $H(z)$ outlined in Eq.~(\ref{H(x)}), 
can also be written as
\begin{equation}
x=p[1-G_1(1-x)]\cdot D[H(p(1-G_0(1-x))] \equiv F(x,p). \label{x_limit}
\end{equation}
For a given $p$, $x$ can be numerically calculated through iteration 
with a proper initial value. 

Correspondingly, using similar arguments, the probability $P_{\infty}(p)$ 
that a randomly chosen node $n$ in the steady state of network $A$ is 
in the giant component is 
\begin{eqnarray} \label{mu1}
\nonumber P_{\infty}(p)&=&p\{\sum_{k}P(k)[1-(1-x)^{k}]\}\times\\
\nonumber &&\sum_{s} h(s)\cdot\{p\sum_{k}P(k)[1-(1-x)^k]\}^{s-1}\\
&=&H(p[1-G_0(1-x)]),
\end{eqnarray}
where $1-(1-x)^k$ is the probability that at least one of the $k$ connectivity
links of node $n$ leads to the giant component. Note that $P_{\infty} (p)$ is also the normalized 
size of the giant component of network $A$ at the steady state.

We find that there is no giant component at the steady state of network 
$A$, i.e., $P_{\infty} (p)=0$ when $p$ is smaller than a critical probability 
$p_c^{II}$ and above the threshold, the giant component appears and
its size increases continuously from $0$ as $p$ increases. This is typical 
second-order phase transition behavior and as $p \rightarrow p_c^{II}$,  
$P_\infty(p_c^{II})=H(p_c^{II}[1-G_0(1-x)]) \rightarrow 0$,  
which suggests $x\rightarrow 0$. Thus we can take the Taylor expansion of  
Eq.~(\ref{x_limit}) with $x\rightarrow 0$ to obtain $ p_c^{II}$ as 
(see Appendix~\ref{ApB}), 
\begin{equation}
p_c^{II}=\frac{1}{Q(0)G_1^{'}(1)}=\frac{\left\langle k \right\rangle}{Q(0)\left\langle k(k-1) \right\rangle}, \label{p_c_2}
\end{equation}
which is consistent with our previous result reported in Ref.\cite{hu2014conditions} 
and depends on $Q(0)$ only but not any other terms from $Q(k_o)$.

In some cases, however, there is no giant component at the steady 
state of network $A$, i.e., $P_{\infty} (p)=0$ when $p$ is smaller 
than a critical probability $p_c^I$ but above the threshold, the giant 
component suddenly appears and its size increases abruptly from $0$ 
as $p$ increases. This is typical first-order phase transition behavior. 
When $p=p_c^I$, the straight line $y=x$ and the curve $y=F(x,p)$ 
from Eq.~(\ref{x_limit}) will tangentially touch each other at $(x_c, x_c$) \cite{hu2014conditions}.
Thus,  the condition corresponding to the first-order transition is that 
the derivatives of both sides of Eq.~(\ref{x_limit}) with respect to $x$ 
are equal, 
\begin{equation}
1=\frac{dF(x,p)}{dx}\vert x=x_c,p=p_c^I. \label{p_c_first}
\end{equation}
Due to the complexity of Eqs.~(\ref{x_limit}) and (\ref{p_c_first}), 
numeric methods are generally used to get $p_c^I$. 

Note $p_c^I=p_c^{II}$ corresponds to the case where the phase transition 
changes from first-order to second-order when the conditions for both 
the first- and second-order transitions are satisfied simultaneously.  
By substituting $p_c^{II}$ from Eq.~(\ref{p_c_2}) into Eq.~(\ref{p_c_first}) 
and further evaluating $x$, we obtain the boundary between the first-order
and second-order phase transitions, which is characterized by (see Appendix \ref{ApC}), 
\begin{equation}
Q(1)=\frac{Q(0)G_1^{''}(1)}{2G_0^{'}(1)}=\frac{Q(0) \left\langle k(k-1)(k-2) \right\rangle}{2 \left\langle k \right\rangle ^2}. \label{1_2_border}
\end{equation}
Thus, the boundary between first- and second-order transitions is 
determined only by the proportion of nodes that do not depend on more than 
one node, i.e., the boundary is solely determined by $Q(0)$ 
and $ Q(1)$ but not any other terms from $Q(k_o)$. This implies that the triple point -- 
the intersection of first order phase transition, second order phase transition 
and the unstable regime is also determined by $Q(0)$ and $ Q(1)$.

When removing any fraction of nodes results in the total collapse of 
network $A$, i.e., when $p_c^{II} \geq 1$, the network is unstable.  
By requiring $p_c^{II}=1$ and using Eq.~(\ref{p_c_2}), we can obtain 
the boundary between the second-order phase transition and the unstable 
state,
\begin{equation}
Q(0)=\frac{1}{G_1^{'}(1)}=\frac{\left\langle k \right\rangle}{\left\langle k(k-1) \right\rangle},\label{2_unstable}
\end{equation}
which depends solely on the proportion of nodes that do not depend on 
other nodes at all, i.e., $Q(0)$.

Similarly, by requiring $p_c^{I}=1$ in Eq.~(\ref{p_c_first}), we use 
numerical calculations to find the boundary between the first-order 
phase transition and the unstable state. Therefore, the complete boundary 
between the unstable state and the phase transition state is achieved 
by joining these two boundaries together. Moreover,  
substituting Eq.~(\ref{1_2_border}) into Eq.~(\ref{2_unstable}), we could obtain the
explicit formula of the triple point which is the intersection of these two boundaries:
\begin{equation}
Q(1)=\frac{\left\langle k(k-1)(k-2) \right\rangle}{2\left\langle k \right\rangle\left\langle k(k-1) \right\rangle}
\end{equation}

Note that for scale-free networks with power law degree distribution $P(k) \varpropto k^{-\gamma}$
and  $\gamma \in (2,3]$, 
both $\left\langle k(k-1) \right\rangle$ and $\left\langle k(k-1)(k-2) \right\rangle$ are divergent. 
This implies that $p_c^{II}=0$ for any $Q(0)$ according to Eq.~(\ref{p_c_2}) 
and the regime of the second order phase transition shrinks towards the 
origin. Thus for scale free networks, the situation becomes 
a little bit simple. Therefore, if $Q(0)>0$ one could always see the second-order 
phase transition with $p_c^{II}=0$ and if $Q(0)=0$ 
the system undergoes unstable or first-order phase transition.

\section{Results on ER Networks}
Section II provided the general framework for random networks with 
an arbitrary degree distribution $P(k)$. We here illustrate it
using an ER network \citep{ER1, ER2,ballobas1985random} with a Poisson 
degree distribution $P(k)=e^{-\left\langle k \right\rangle}{\left\langle k \right\rangle}^k/{k!}$
where $\left\langle k \right\rangle$ is the average degree. We choose 
this network because it is representative of random networks, and the 
generating function corresponding to the degree distribution $P(k)$ is
$G_0(z)=e^{\left\langle k \right\rangle(z-1)}$. 

\subsection{Second-order phase transitions}
\begin{figure}
\includegraphics[width=0.4\textwidth, angle = 0]{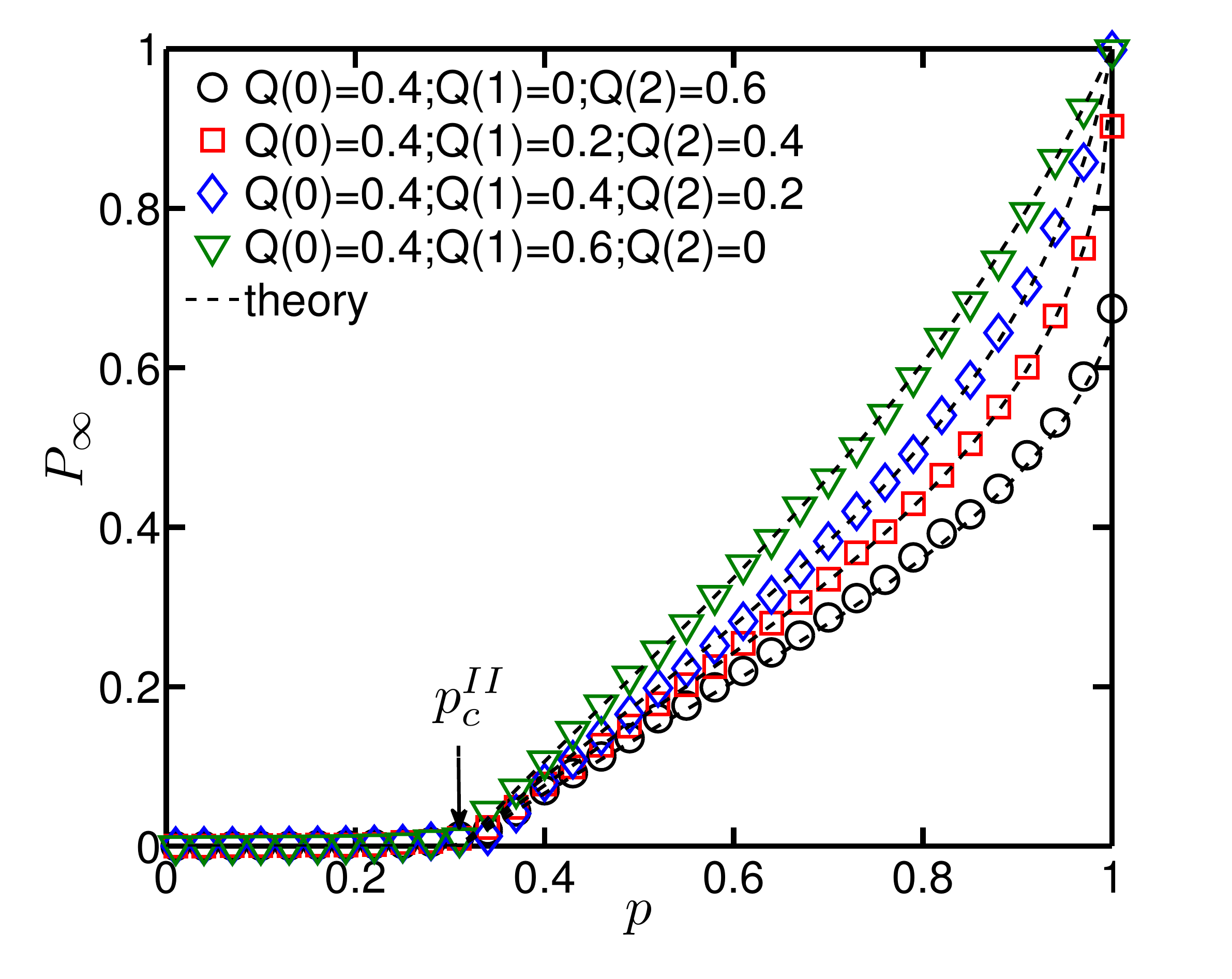}
\caption{\label{Fig2}(Color online) The size of the giant component $P_\infty(p)$,
as a function of the fraction of nodes that remain after random removal, $p$, 
for ER networks with $\left\langle k \right\rangle =8$ and $D(z)=Q(0)+Q(1)z+Q(2)z^2$. 
The symbols represent simulation results of $10^4$ nodes and the dashed lines 
show the theoretical predictions from Eq.~(\ref{mu1}). The percolation threshold $p_c^{II}$ 
is uniquely determined by $Q(0)$.}
\end{figure}

 Plugging $G_1^{'}(1)=\left\langle k \right\rangle$ into Eq.~(\ref{p_c_2}),  we get the second-order 
 phase transition point $p_c^{II}$,
\begin{equation}
p_c^{II}=\frac{1}{Q(0)\left\langle k \right\rangle }.  \label{ER_2}
\end{equation}
Therefore, for ER networks, the critical point of second-order phase 
transition is indeed determined solely by $Q(0)$ and its average degree.  
We support our analytical results by simulations. We choose $\left\langle k \right\rangle=8$  
and $D(z)=Q(0)+Q(1)z+Q(2)z^2$ with $Q(0)$ fixed at 0.4 and 
$Q(1)$, $Q(2)$ tunable.  Fig.~\ref{Fig2} shows the size of the 
giant component $P_\infty(p)$ as a function of $p$ with the given $\left\langle k \right\rangle$
and $D(z)$.  Note that in all cases simulation results (symbols) agree 
well with numerical results (dotted lines) and the curves of $P_{\infty}(p)$ 
converge at a fixed value of $p_c^{II}=0.3125$ as predicted by Eq.~(\ref{ER_2}). 
This convergence of $P_{\infty}(p)$ curves is possible because $p_c^{II}$ is 
determined solely by $Q(0)$, which is fixed to be 0.4 in Fig.~\ref{Fig2}.
Note that if there is no directed dependency links in the network, i.e., $Q(0)=1$,  
we will get $p_c^{II}=1/\left\langle k \right\rangle$, which is consistent
with the well-known result obtained in Ref.~\cite{cohen2000resilience}.

\subsection{First-order phase transitions}
When networks have a greater proportion of directed dependency links,  
an abrupt transition can occur instead of a continuous transition demonstrated 
in Fig.~\ref{Fig2}. To get the $p_c^{I}$ for the onset of this abrupt transition, 
we equate the derivatives of both sides of Eq.~(\ref{x_limit}) with respect to $x$, i.e.,
\begin{equation}
1=\frac{d\{p(1-e^{-\left\langle k \right\rangle x})\cdot D[H(p(1-e^{-\left\langle k \right\rangle x})\}}
{dx}\vert_{x=x_c,p=p_c^{I}}, \label{p_c_1}
\end{equation}
where we used the equtions $G_0(z)=G_1(z)=e^{\left\langle k \right\rangle(z-1)}$.
Using Eqs.~(\ref{x_limit}) and (\ref{p_c_1}),  we apply numerical methods to get $p_c^I$. 

\begin{figure}
\includegraphics[width=0.4\textwidth, angle = 0]{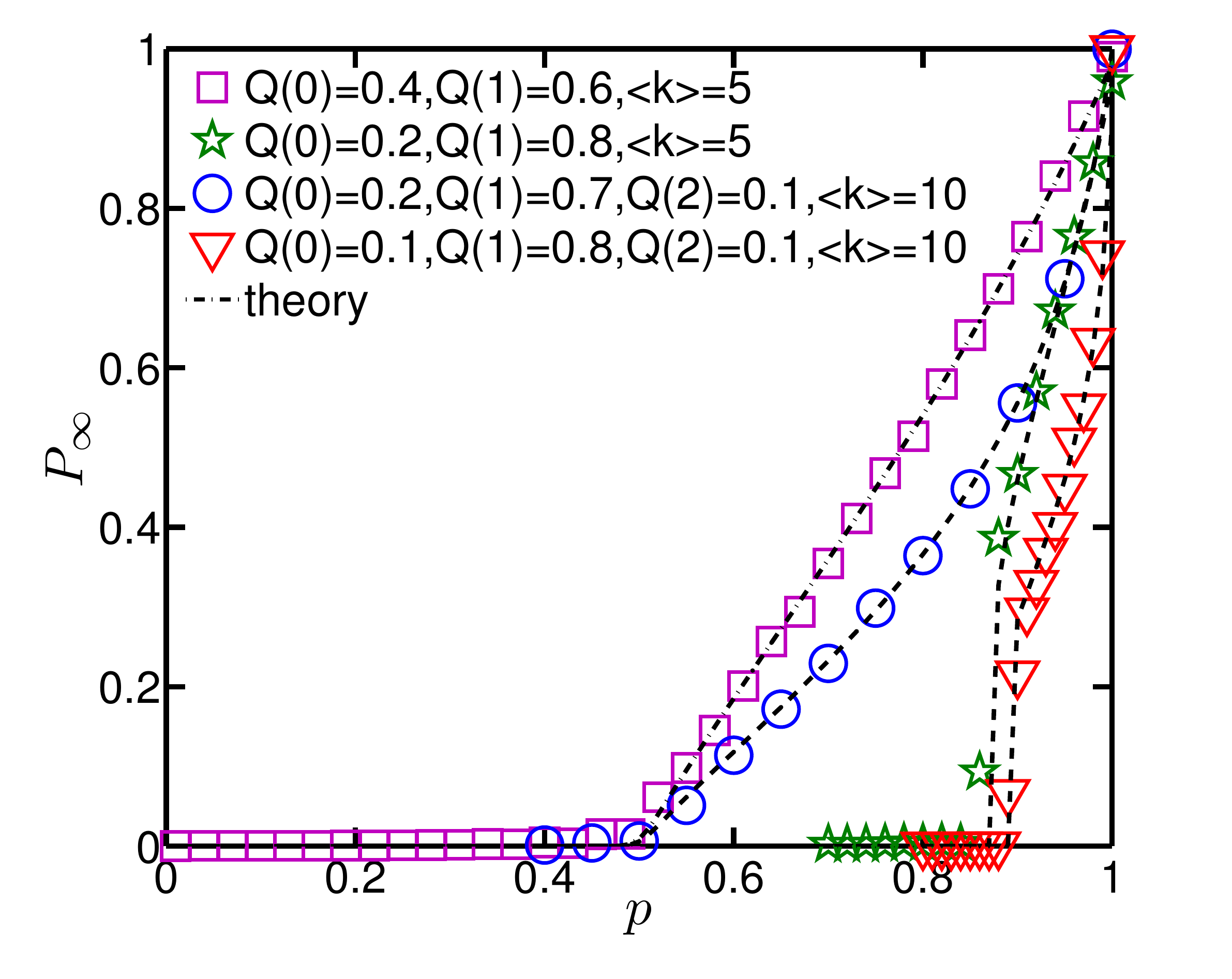}
\caption{\label{Fig3}(Color Online) The size of the giant component 
$P_\infty$ as a function of the fraction of nodes that remain after 
random removal, $p$, for ER networks. Here we used $D(z)=Q(0)+Q(1)z$ 
with $\left\langle k \right\rangle =5$ ($\square$ and $\star$ ) and 
$D(z)=Q(0)+Q(1)z+Q(2)z^2$ with $\left\langle k \right\rangle=10$ ($\bigcirc$ 
and $\triangledown$). The symbols represent simulation results of $10^4$ 
nodes and the dashed lines are the theoretical predictions from Eq.~(\ref{mu1}). 
With a relatively larger $Q(0)$, the network undergoes a second-order phase 
transition at $p_c^{II}$, which only depends on $Q(0)$. However for relatively 
smaller $Q(0)$ and larger $Q(1)$ and $Q(2)$, the network undergoes a 
first-order phase transition.}
\end{figure}

With $D(z)=Q(0)+Q(1)z+Q(2)z^2$,  Fig.~\ref{Fig3} shows the size of
the giant component $P_\infty(p)$ as a function of $p$ by comparing 
simulation results and theoretical predictions. Note that they agree with 
each other very well. Fig.~\ref{Fig3} shows that with $\left\langle k \right\rangle=5$ 
and $Q(0)+Q(1)=1$, when $Q(0)=0.4$, $P_\infty(p)$ undergoes a 
second-order phase transition at $p_c^{II}=0.5$ ($\square$), but when 
$Q(0)=0.2$, $P_\infty(p)$ exhibits behavior of a first-order 
phase transition at $p_c^{I}$, satisfying Eq.~(\ref{p_c_1}) ($\bigstar$). 
In addition, when $\left\langle k \right\rangle=10$, $Q(0)=0.2$, $Q(1)=0.7$ 
and $Q(2)=0.1$, $P_\infty(p)$ undergoes a second-order phase transition  
at $p_c^{II}=0.5$ ($\bigcirc$), but when $Q(0)=0.1$, $Q(1)=0.8$ and $Q(2)=0.1$, 
$P_\infty(p)$ undergoes a first-order phase transition at $p_c^{I}$ 
predicted by Eq.~(\ref{p_c_1}) ($\triangledown$).
\subsection{Boundaries of phase diagram}
We fix the average degree $\left\langle k \right\rangle$ and from Eq.~(\ref{ER_2})
we conclude that the smaller $Q(0)$ in the network, the bigger the $p_c^{II}$ value. 
If $Q(0)$ is properly small that $p_c^{II} \approx1$, which corresponds to the case 
in which the removal of any fraction of nodes causes a second-order phase transition
that totally disintegrates network $A$. Thus, by requiring $p_c^{II}=1$, and using 
Eq.~(\ref{ER_2}) we obtain the boundary between the second-order phase transition 
and the unstable state,
\begin{equation}
\frac{1}{\left\langle k \right\rangle Q(0)}=1. \label{Unstable}
\end{equation}
In addition, using Eq.~(\ref{1_2_border}), we obtain the boundary between the first-order 
and second-order phase transitions of network $A$,
\begin{equation}
Q(1)=\frac{\left\langle k \right\rangle Q(0)}{2}. \label{Border12}
\end{equation}

\begin{figure}
\includegraphics[width=0.4\textwidth, angle = 0]{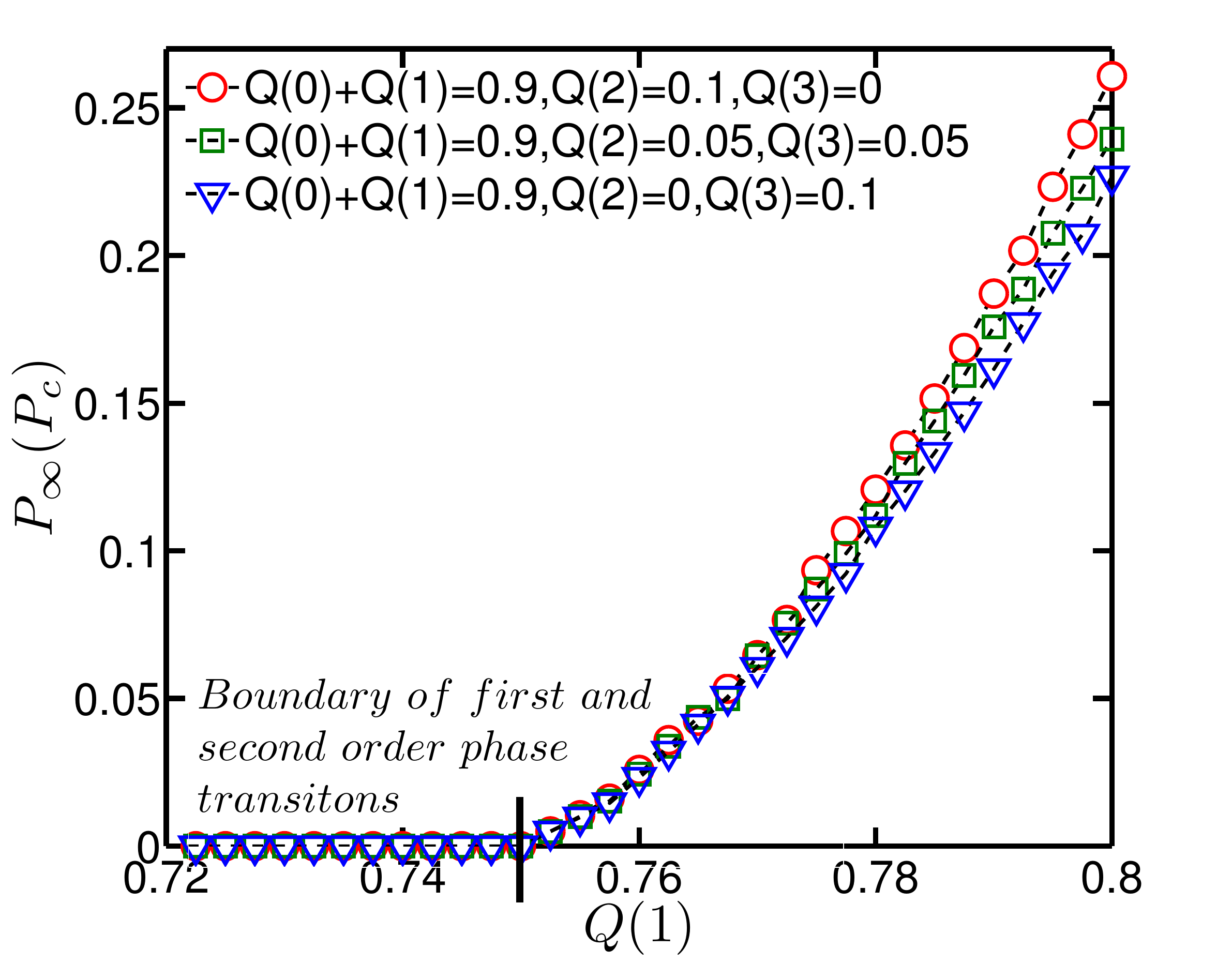}
\caption{\label{Fig4} (Color online) Comparison between simulation (symbols) and 
theory (lines) for $P_\infty(p_c)$ as a function of $Q(1)$ for different $D(z)$ 
($D(z)=Q(0)+Q(1)z+Q(2)z^2$) while keeping $Q(0)+Q(1)=0.9$ and $\left\langle k \right\rangle=10$. 
At the first-order phase transition point $p_c^I$, $P_\infty(p_c)$ is nonzero; 
whereas at the second-order phase transition point $p_c^{II}$ and $P_\infty(p_c)$ is zero.
From  Eq.~(\ref{Border12}) the boundary between first-order phase transition 
and second-order phase transition is only dependent on $Q(0)+Q(1)$, 
thus at $Q(1)_c=0.75$ for this case. }
\end{figure}

Using $D(z)=Q(0)+Q(1)z+Q(2)z^2+Q(3)z^3$ where $Q(0)+Q(1)=0.9$ 
and $\left\langle k \right\rangle=10$, Fig.~\ref{Fig4} plots $P_\infty(p_c)$ 
as a function of $Q(1)$ by comparing simulation and numerical results.
The critical value of $Q(1)_c$ falls onto $Q(1)_c=0.75$ as predicted by Eq.~(\ref{Border12}), 
delimiting two different transition regimes. Specifically,  if $Q(1)<0.75$,  
$P_\infty(p_c)=0$, which indicates the presence of a second-order phase transition,  
but if $Q(1)>0.75$, $P_\infty(p_c)>0$, which indicates the presence of a first-order 
phase transition.
We also consider a special case in which $Q(0)+Q(1)=M$ and use Eq.~(\ref{Border12}) 
to determine the boundary between the first-order phase transition and  the second-order 
phase transition,
\begin{equation}
Q(1)=\frac{M\left\langle k \right\rangle}{\left\langle k \right\rangle+2}. \label{q_1_1}
\end{equation}
In addition, in terms of $M$,  Eq.~(\ref{Unstable}) delivers the boundary  between 
the second-order phase transition and unstable state,
\begin{equation}
Q(1)=\frac{M\left\langle k \right\rangle -1}{\left\langle k \right\rangle}. \label{q_1_2}
\end{equation}
\begin{figure}
\includegraphics[width=0.39\textwidth, angle = 0]{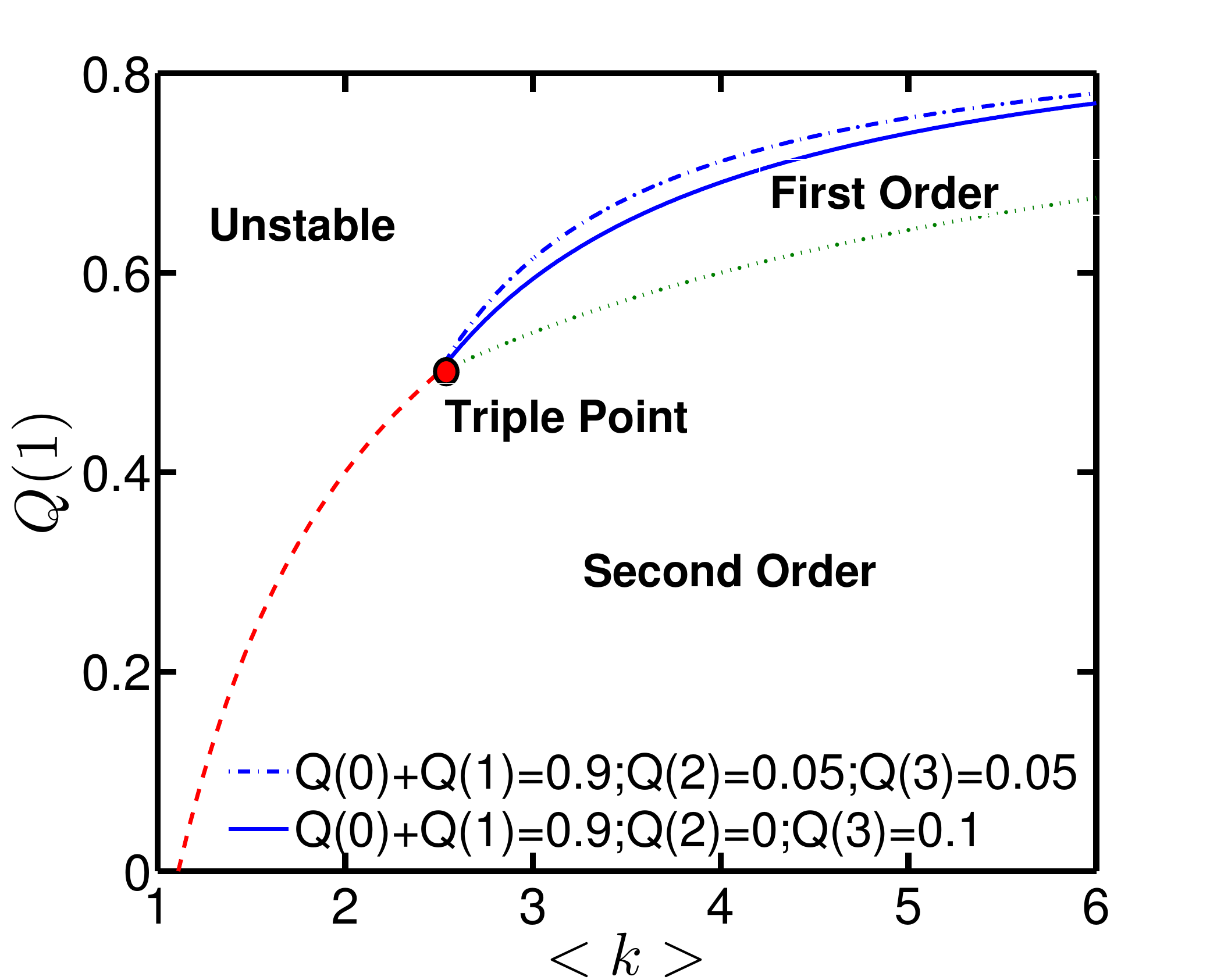}
\caption{\label{Fig5} (Color online) The fraction of nodes that have one dependent 
node $Q(1)$ as a function of the average degree $\left\langle k \right\rangle$ with 
$Q(0)+Q(1)=\frac{9}{10}$. The dashed lines are theoretical results obtained from
 Eqs.~(\ref{q_1_1}) (green) and (\ref{q_1_2}) (red) with intersection points at $(\frac{2}{2M-1},\frac{1}{2})$.  
 The dashed blue line is the boundary between first-order phase transition and unstable system, 
 obtained numerically. Here the dashed red and green lines only depend on $m_0$ whereas 
 the blue lines (both solid and dashed) depend on the specific details of $Q(k_o)$ other than $M$.  }
\end{figure}
Thus, in the coordinate system of $\left\langle k \right\rangle$-$Q(1)$, 
using Eqs.~(\ref{q_1_1}) and (\ref{q_1_2}) we can plot the phase diagram 
of network $A$ under random failures, with these two boundaries converging 
at the triple point $(\frac{2}{2M-1},\frac{1}{2})$ (the solid
red dot in Fig.~\ref{Fig5}). Because $\left\langle k \right\rangle >0$ 
always holds, when $M\leq \frac{1}{2}$ this intersection point is non-physical, 
indicating that the network will not be subject to first-order phase transitions 
under attack irregardless of the form of $P(k)$, but if $M> \frac{1}{2}$, 
the network will be subject to first-order phase transitions. 

Fig.~\ref{Fig5} shows the boundaries in the phase diagram with 
$M=\frac{9}{10}>\frac{1}{2}$, where the boundaries between 
first-order phase transitions and the unstable state are determined 
numerically. Note that,  when $M$ is fixed, the boundary between  
the second-order phase transition and the unstable state (dashed red line) 
as well as the boundary between the first-order and second-order phase 
transitions (dashed green line) are also fixed because they depend only on $M$,
but the boundary between the first-order phase transition and the unstable state 
(dashed blue line) is subject to the details of $Q(k_o)$. For example, when 
$Q(0)+Q(1)=\frac{9}{10}$, a shuffle of the remaining terms in $Q(k_o)$ causes
a shift in the boundary line, shown as the displacement of the solid blue line
to the dashed blue line in Fig.~\ref{Fig5}.

\section{CONCLUSIONS}
In summary, we present an analytical formalism for studying random networks 
with both connectivity links and directed dependency links under random node failures. 
Using a probabilistic approach, we find that the directed dependency links greatly 
reduce the robustness of a network. We show that the system disintegrates in a 
form of second-order phase transition at a critical threshold and the boundary 
between second-order phase transition and unstable regimes solely determined 
by the proportion of nodes that do not depend on other nodes. Our framework also provides 
the solution for the boundary between the first-order and second-order phase transitions, 
which is characterized by the proportion of nodes that depend on no more than one node. 
\section*{Acknowledgments}
This work is partially supported by the NSFC grant no. 61203156.
The Boston University work is supported by NSF grant no. CMMI 1125290 and
DTRA grant no. HDTRA1-14-1-0017.

\appendix

\section{}\label{ApB}
If $p \rightarrow p_c^{II}$,  $x\rightarrow 0$. From Eq.~(\ref{x_limit}) we have
\begin{equation} \label{b1}
1-G_1(1-x)=G_1^{'}(1)x-\frac{G_1^{''}(1)}{2}x^2+ O(x^3),
\end{equation} 
\begin{equation}\label{b2}
1-G_0(1-x)=G_0^{'}(1)x-\frac{G_0{''}(1)}{2}x^2+O(x^3),
\end{equation}
and
\begin{eqnarray}\label{b3}
\nonumber D\{H\{p[1-G_0(1-x)]\}\}&=&Q(0)+pQ(1)h(1)G_0^{'}(1)x \\
&&+O(x^2).
\end{eqnarray} 
Using Eqs.~(\ref{b1}), (\ref{b2}) and (\ref{b3}), we can write Eq.~(\ref{x_limit}) as 
\begin{eqnarray} \label{b4}
\nonumber x&=&pQ(0)G_1^{'}(1)x+p[ pQ(1)h(1) G_0^{'}(1)G_1^{'}(1) \\
&& -\frac{Q(0) G_1^{''}(1)}{2}] x^2+O(x^3).
\end{eqnarray}
Since $x\in (0,1)$, we can divide both sides of Eq.~(\ref{b4}) by $x$, and obtain 
\begin{eqnarray} \label{b5}
\nonumber 1&=&pQ(0)G_1^{'}(1)+p[pQ(1)h(1)G_0^{'}(1)G_1^{'}(1)  \\
&& -\frac{Q(0) G_1^{''}(1)}{2}]x+O(x^2).
\end{eqnarray}
As $x \rightarrow 0$,  taking the limits of both sides of Eq.~(\ref{b5}) we get 
\begin{equation}\label{b6}
p_c^{II}=\frac{1}{Q(0)G_1^{'}(1)}.
\end{equation}
\section{} \label{ApC}
Putting Eq.~(\ref{b6}) back into Eq.~(\ref{b5}), we get
\begin{equation}\label{c1}
\frac{ Q(1)h(1)G_0^{'}(1)}{Q(0)}x=\frac{Q(0)G_1^{''}(1)}{2}x+ O(x^2).
\end{equation}
To simplify Eq.~(\ref{c1}), we first take the derivatives of both sides of Eq.~(\ref{H(x)}) 
with respect to $x$ and obtain
\begin{equation}\label{c2}
H^{'}(z)=D(H(z))-z \frac{\partial (D(H(z)))}{\partial (H(z))}H^{'}(z).
\end{equation}
Plugging $z=0$ into Eq.~(\ref{c2}), we get $H^{'}(0)=D(H(0))=D(0)=Q(0)$. 
Using Eq.~(\ref{H}), we easily obtain $H^{'}(0)=h(1)$ and thus
 $h(1)=Q(0)$, which would reduce Eq.~(\ref{c1}) as
\begin{equation} \label{c4}
Q(1)x=\frac{Q(0)G_1^{''}(1)}{2G_0^{'}(1)}x +O(x^2).
\end{equation}

Up to this point, if $x\rightarrow x_t=0$, network $A$ undergoes a second-order phase transition  
and thus Eq.~(\ref{c4}) clearly holds, but if $x\rightarrow x_t \neq 0$, network $A$ undergoes
a first-order phase transition.  On the boundary between the first-order 
and the second-order phase transitions,  we get a nonzero $x_t$, but it is negligibly small. Here, 
we can treat $O(x_t) \approx 0$ and obtain the condition characterizing this boundary as
\begin{equation}\label{c5}
Q(1)=\frac{Q(0)G_1^{''}(1)}{2G_0^{'}(1)}.
\end{equation} 

\bibliographystyle{apsrev}
\bibliography{References}
\end{document}